\documentclass[12pt]{article}
\usepackage{amsfonts,amsthm,amsmath,amssymb}
\usepackage{hyperref}
\usepackage{cite}
\usepackage[paper=letterpaper,margin=1in]{geometry}
\usepackage{graphicx}
\usepackage{units}

\def\K{\mathcal{K}}
\def\t{\tau}
\def\a'{\alpha'}

\renewcommand\bar{\overline}

\newcommand{\be}{\begin{equation}}
\newcommand{\ee}{\end{equation}}
\newcommand{\bea}{\begin{eqnarray}}
\newcommand{\eea}{\end{eqnarray}}
\newcommand{\ba}{\begin{array}}
\newcommand{\ea}{\end{array}}
\newcommand{\ben}{\begin{enumerate}}
\newcommand{\een}{\end{enumerate}}
\newcommand{\bi}{\begin{itemize}}
\newcommand{\ei}{\end{itemize}}
\newcommand{\bc}{\begin{center}}
\newcommand{\ec}{\end{center}}
\newcommand{\bfig}{\begin{figure}}
\newcommand{\efig}{\end{figure}}
\newcommand{\nn}{\nonumber}
\newcommand{\eref}[1]{(\ref{#1})}

\numberwithin{equation}{section}

\begin{document}
\thispagestyle{empty}
\renewcommand{\thefootnote}{\fnsymbol{footnote}}

\begin{flushright}
\end{flushright}
\vspace{1 cm}

\begin{center}

\bf{{\LARGE Natural Inflation from Near Alignment\\ in Heterotic String Theory}\\

\vspace{1cm}}

{Tibra Ali$^{a,}$\,\footnote[4]{\tt{tibra.ali@pitp.ca}}, S. Shajidul Haque$^{b,}$\,\footnote[3]{{\tt{shajid.haque@wits.ac.za}}}, Vishnu Jejjala$^{b,}$\,\footnote[1]{\tt{vishnu@neo.phys.wits.ac.za}}}\\

\vspace{0.5 cm}\vspace{0.2 cm}{{\it$^{a}$Perimeter Institute for Theoretical Physics \\31 Caroline Street N., Waterloo, ON N2L 2Y5, Canada} }

\vspace{0.5 cm}{{\it$^{b}$National Institute for Theoretical Physics, School of Physics,\\ and Mandelstam Institute for Theoretical Physics,\\University of the Witwatersrand, Johannesburg, WITS 2050, South Africa }} \\

\vspace{2cm}
{\bf Abstract}
\end{center}
\begin{quotation}
\noindent
We propose a model for large field inflation in heterotic string theory.
The construction applies the near alignment mechanism of Kim, Nilles, and Peloso.
By including gaugino condensates and world-sheet instanton non-perturbative effects, we obtain a large effective axion decay constant.
\end{quotation}

\setcounter{page}{0}
\setcounter{tocdepth}{2}
\newpage
\section{Introduction}
\renewcommand{\thefootnote}{\arabic{footnote}}
\setcounter{footnote}{0}
Cosmological inflation~\cite{Guth:1980zm} uses a period of exponential expansion immediately after the Big Bang to account for why the Universe appears homogeneous, isotropic, and flat on large scales today.
Scalar field models of inflation moreover require that the inflationary potential $V(\varphi)$ is extremely flat~\cite{Linde:1981mu, albrecht1982cosmology}.
This potential is characterized by the slow-roll parameters:
\be
\varepsilon\sim \frac12 M_\mathrm{Pl}^2 \left( \frac{V'(\varphi)}{V(\varphi)} \right)^2 \ll 1 ~, \qquad
|\eta| \sim M_\mathrm{Pl}^2 \left| \frac{V''(\varphi)}{V(\varphi)} \right| \ll 1 ~, \label{eq:slowroll}
\ee
where $M_\mathrm{Pl}$ is the reduced Planck mass.
Current astrophysical data are beginning to illuminate the structure of the potential.
The Planck satellite's measurement of the spectral index~\cite{Planck} establishes that $\varepsilon$ or $\eta$ are of order $10^{-2}$ and favours power law potentials with leading exponent less than two.
Intriguing preliminary results on B-modes from the BICEP2 collaboration~\cite{Ade} suggest that the energy scale for inflation is set by the tensor-to-scalar ratio $r$ as follows:
\be
V \simeq \left(\frac{r}{0.2}\right) \times \left( 2.2\times 10^{16}\ \mathrm{GeV} \right)^4 ~.
\ee
However, dust within the galaxy contaminates the BICEP2 measurements at the same level as the signal~\cite{Planck1} and therefore the observations in~\cite{Ade} should be regarded as a provisional hypothesis that awaits refinement through improved analysis.

Since the Lyth bound~\cite{lyth, bg} indicates that the motion $\Delta\varphi$ in field space is
\be
\Delta\varphi \simeq \sqrt{\frac{r}{0.01}}\ M_\mathrm{Pl} ~,
\ee
the still inconclusive BICEP2 claim that $r\simeq 0.2$ advocates a trans-Planckian fluctuation in the inflaton field $\varphi$ over the course of inflation, while at the same time retaining the form of the potential implied by~\eref{eq:slowroll}.
Since $M_\mathrm{Pl}$ is the ultraviolet cutoff, trans-Planckian motion in field space raises questions about the validity of the effective field theory description of the physics.
From the Wilsonian perspective, the flat potential ought to be destabilized by higher dimensional contributions to the effective action for the scalar.
Indeed, we may write the effective action in terms of a Lagrangian density
\be
{\cal L} =- \frac12 \partial_\mu \varphi \partial^\mu \varphi - \left[ V_0 + \frac12 m^2 \varphi^2 + \lambda \varphi^4 + \sum_{k=1}^\infty c_k \left( \frac{\varphi^{4+k}}{M_\mathrm{Pl}^k} \right) \right] ~,
\ee
where the $c_k$ are generically order one coefficients, and the terms in the sum become problematic for $\Delta\varphi \gtrsim M_\mathrm{Pl}$.

One scenario that yields a flat potential is so-called \textit{natural inflation}~\cite{Freese}, in which an axion serves as the inflaton.
The existence of a shift symmetry under which the potential is invariant ensures its flatness.
This symmetry persists (at least approximately) non-perturbatively at the quantum level.
Because the setup enforces symmetries on the ultraviolet completion, realizing the model within the architecture of a fundamental theory of quantum gravity such as string theory is remarkably challenging~\cite{bdfg,ArkaniHamed}.
Large field models such as N-flation~\cite{dkmw} or axion monodromy~\cite{silverstein1, Silverstein2, knl, kal} are supported in this scenario.

Kim, Nilles, and Peloso (KNP)~\cite{kim} propose a framework for large field models in which two axionic fields are balanced.\footnote{
The idea of multi-field assisted inflation was earlier studied in~\cite{Liddle:1998jc}.}
The alignment mechanism advanced by KNP generates a large effective axion scale even though the axion decay constants are themselves sub-Planckian.
Recently~\cite{long} constructed a realization of this alignment mechanism in type IIB superstring theory.
Inspired by~\cite{long}, we develop in this article the ideas of KNP~\cite{kim} (further explained in~\cite{Kappl:2014lra}) in the context of the $E_8\times E_8$ heterotic string.
While KNP commented on the possibility of an embedding~\cite{Ibanez} in heterotic strings in their paper~\cite{kim}, the purpose of our work is to embed the KNP mechanism in a rather different and strikingly simple heterotic scenario.
The question of moduli stabilization is well developed within this framework~\cite{Anderson}.

Though the ten dimensional critical string theories are related by dualities, the construction in heterotic theory is notably different from~\cite{long}.
Similar recent attempts to embed natural inflation in type IIB string theory or string motivated supergravity can be found in~\cite{Czerny:2014qqa,Gao:2014uha,Kenton:2014gma,Ashoorioon:2009wa}.

While most investigations of cosmology in string theory are couched in the language of the type II superstring, the heterotic string has considerable advantages of its own.
In particular, modeling cosmological inflation is only a part of a larger story.
A resolution to the vacuum selection problem must also justify the existence of the Standard Model at low energies.
Since~\cite{chsw}, the heterotic string has offered a promising avenue toward this goal.\footnote{
Of course, there is not a uniform opinion on how to arrive at the Standard Model from a theory of strings and D-branes, and numerous alternatives to heterotic compactification have been considered.
See, for example,~\cite{iu}.}
Several realistic constructions of a theory of particle physics with three generations of chiral matter that fall in representations of $SU(3)_C\times SU(2)_L\times U(1)_Y$ with the appropriate field interactions stem from Calabi--Yau compactifications of the heterotic string~\cite{Braun1, Braun3, br, bmw, Anderson4,  Anderson5}.
Moreover, within the heterotic setup, one can stabilize geometric moduli for the compactification~\cite{Anderson2, Anderson, Anderson3}.
Engineering a stable, aligned, natural inflation scenario within heterotic string theory is therefore a step toward synthesizing model building in cosmology with model building in particle physics.
Since the two endeavours are today mostly disparate enterprises, facilitating their marriage serves as a principal motivation for our work.

The organization of the paper is as follows.
In Section~\ref{sec:niam}, we briefly review natural inflation and the decay constant alignment mechanism.
In Section~\ref{sec:model}, we construct a model in which a pair of axion fields corresponds to K\"ahler moduli.
The scalar potential within this construction has a large effective decay constant as a consequence of near alignment.
In Section~\ref{sec:example}, we provide a numerical example with order one numbers to illustrate the naturalness of the construction.
In Section~\ref{sec:disc}, we discuss the results and indicate future work.
The Appendix derives the form of the scalar potential.

\section{Natural Inflation and the Alignment Mechanism}\label{sec:niam}
One way to guarantee a small value for the $\eta$ parameter is when the inflaton field is protected by a global symmetry.
If inflation is driven by an axion $\tau$, then this is easy to achieve as the axion has a continuous shift symmetry, $\tau\to \tau+2 \pi f$, where $f$ is the axion decay constant.
However, non-perturbative corrections break this continuous symmetry to a discrete shift symmetry.
An example of a potential with such a shift symmetry is
\begin{align}
V(\tau)=\Lambda^4 \left[1\pm\text {cos} \left (\frac{\tau}{f} \right)\right] ~. \label{eq:one-field}
\end{align}
For large values of $f$, this potential supports inflation.
The higher order terms are suppressed by $f$, and this ensures a small value for the slow-roll parameter $\eta$.

In principle, it is difficult to generate such potentials with decay constants $f$ larger than the Planck scale from string theory.
Typically, instanton corrections spoil the flatness of the potential~\cite{bdfg,ArkaniHamed}.
The KNP proposal~\cite{kim} is useful in this respect.
It shows that if there is more than one axion with multiple non-perturbative terms in the potential generated for different linear combinations of these axions, then a large effective decay constant can be generated from one of the linear combinations.
The other linear combinations have much smaller effective decay constants that produce high curvatures of the potential in those directions.
These other linear combinations are frozen to some value by the potential.

For example, suppose, for two axions $\tau_1$ and $\tau_2$, we have a potential that has the following form:
\begin{align}
V = \Lambda^4_A \left[1- \cos{\left(\frac{\tau_1}{f_{A_1}}+\frac{\tau_2}{f_{A_2}}\right)}\right] + \Lambda^4_B \left[1- \cos{\left(\frac{\tau_1}{f_{B_1}}+\frac{\tau_2}{f_{B_2}}\right)}\right] ~. \label{eq:original_potential}
\end{align}
It might happen that the decay constants in this potential are \emph{aligned}, \textit{i.e.}, they satisfy the following condition:
\begin{align}
\frac{f_{A_1}}{f_{A_2}}=\frac{f_{B_1}}{f_{B_2}} ~.
\end{align}
If so, the same linear combination of axions appears in both of the cosine functions, and an orthogonal direction is the flat direction.
It may happen that this alignment is slightly broken in a natural way, and then we get a nearly flat direction that can have a large \emph{effective} decay constant.
This supplies an inflaton field.
Crucially, the parameters $f$ are sub-Planckian, and therefore the effective field theory description of the system is valid.
\section{The Model}\label{sec:model}
We start with $E_8\times E_8$ heterotic string theory compactified on a Calabi--Yau threefold $X$.
We shall assume that compactification on $X$ gives rise to an effective $S, T^1, T^2, Z^a$ model.
We denote by $T^i$ the complex moduli fields coming from the K\"ahler structure deformations while $Z^a$ denotes the moduli fields from the complex structure deformations.
Here, we assume that $X$ is chosen in such a way that  there are two K\"ahler moduli fields but we shall not need to specify the number of fields coming from the complex structure deformations.
We also have the universal axiodilaton field which we denote by $S$.
It is important to note that, instead of adopting the standard embedding, we shall loosely adopt the setting outlined in~\cite{Anderson, Anderson2, Anderson3}, in which moduli fixing is achieved by making a judicious choice of the gauge bundle and non-perturbative superpotentials coming from gaugino condensates and string instantons.

As mentioned previously, our construction shares some similarities with the type IIB construction advanced by ~\cite{long}.
Thus, we require the interplay between the classical superpotential and non-perturbative contributions.
More specifically we shall need the classical superpotential to depend entirely on $Z^a$ moduli.
We also require \emph{two} sources of non-perturbative potentials for the K\"ahler moduli, which we achieve by turning on gaugino condensations as well as string instantons.
We shall assume that using the methods outlined in~\cite{Anderson} and related works that we can fix all the complex structure moduli fields as well as the real parts of $T^1$ and $T^2$.
We as well assume that we can fix $S$ to a value that keeps us in the perturbative regime by imposing the F-term equation involving non-perturbative gaugino condensations~\cite{Cicoli}.
While we shall not make very precise assumptions for the values of $Z^a$, we shall pick explicit values for the real parts of $T^1, T^2$ and $S$ in our construction.
We then tune the D-term proposed in~\cite{Anderson} and generate an effective potential of the form (\ref{eq:original_potential}) for the remaining two axion fields $\tau_1$ and $\tau_2$ coming from the two K\"ahler structure moduli fields.

\subsection*{Superpotential and K\"ahler Potential}
As we shall be using the axions from the K\"ahler sector for decay constant misalignment, it will be necessary for the perturbative superpotential $W_p$ to be independent of the K\"ahler sector moduli.
Apart from the fact that $W_p$ has to be small but non-zero, our construction is independent of the details of how that is achieved.
In~\cite{Anderson}, an F-term classical superpotential was used for $Z^a$ to stabilize to a supersymmetric Minkowski ground state.
At this ground state their $W_p$ is zero.
However, upon closer inspection it becomes clear that this is not the most general solution (for example, the bundle moduli are all set to zero) and we assume that it is possible to modify their proposal in a way such that the stabilized values of $Z^a$ yield a small but non-zero value for $W_p$.\footnote{
Alternatively one can envision a scenario where fractional Wilson lines are used to generate a small $W_p$ as in~\cite{Gukov}.
We are grateful to L.\ McAllister for bringing this work to our attention.}
Thus we choose the perturbative superpotential to be exclusively a function of the complex structure moduli:
\begin{equation}
W_p=W_p(Z^a) ~. \label{eq:0th-superpotential}
\end{equation}
For definiteness, we take a K\"ahler potential:
\begin{align}
K=-\log{(S+\bar{S})} -  \log{8\K} -  \log{8\widetilde{\K}} ~, \label{eq:Kahler}
\end{align}
where $S$ is the axiodilaton, and we decompose the axiodilaton in terms of real scalar fields:
\begin{align}
S=s+i\sigma.
\end{align}
$\K$ and $\widetilde{\K}$ are the K\"ahler potentials in the complex structure and the K\"ahler structure sectors, respectively.
For this particular form of the superpotential we assume that the contribution from the complex structure moduli is such that $e^{-  \ln{8\K}} \sim 1$, and we therefore neglect these effects in the potential that we calculate below.

In the large volume limit the quantum corrections to $\widetilde{\K}$ are small, and it takes the form
\begin{align}
\widetilde{\K} = \frac{1}{8}d_{ijk} (T^i+\bar{T}^i)(T^j+\bar{T}^j)(T^k+\bar{T}^k) ~,
\end{align}
where $d_{ijk}$ are triple intersection numbers of the Calabi--Yau manifold $X$.

We decompose the two K\"ahler structure moduli fields in terms of their component fields as 
\begin{align}
T^1 &= t_1 + i \tau_1 ~,\nonumber\\
T^2 &= t_2 + i \tau_2 ~. \label{eq:Kahler-fields}
\end{align}
To emphasize the key points, we consider the case that all the intersection numbers except for $d_{111}=1$ and $d_{222}=-1$ are zero.
This is certainly not a necessary assumption, but it will simplify our computations significantly and demonstrate the idea within a simple setup.

\subsection*{Non-perturbative Terms}
To generate the desired potential for $\t^i$ we shall need to turn on non-perturbative terms for $T^i$ coming from gaugino condensates of the gauge line bundles as well worldsheet instantons. By choosing our gauge bundle to be of the form $V=\mathcal{U}\oplus_I \mathcal{L}_I$ where $\mathcal{L}_I$ denotes $U(1)$ line bundles, we generate non-perturbative gaugino condensate terms~\cite{Anderson, Cicoli} in the superpotential of the form
\begin{align}
W^{\mathrm{gaugino}}_{\mathrm{np}} = A' e^{-\alpha(S-\beta_i T^i)} ~, \label{eq:non-pert-1}
\end{align}
where $A'$ is a constant and $\beta_i$ and $\alpha$ are defined below. In our case $S=f$, where $f$ is the four-dimensional gauge kinetic function.
We fix the normalization of $S$ by setting its real part $s=\frac{8\pi}{g^2_4}$. We can then use the above superpotential and $W_p$ to fix the value of the moduli $s$ to some value $s^*$ by imposing the $F$-term equation for $S$. We then absorb the $s^*$ dependence into the constant prefactor.
The constant $\alpha=\frac{3\pi}{b}$ where $b$ is the coefficient of the one-loop beta-function. For $E_8$ we have $b=90$, yielding $\alpha=\frac{\pi}{30}$.
$\beta_i$ are given by
\begin{align}
\beta_i = \int_X (\mathrm{ch}_2(V)-\frac{1}{2}\mathrm{ch}_2(TX))\wedge \omega_i ~,
\end{align}
where $\mathrm{ch}_2$ is the second Chern class of the bundle indicated in its argument and $TX$ is the tangent bundle of $X$. The $\omega_i$ supply a basis of the second cohomology class.
We shall also need non-perturbative contributions from another sector, namely the worldsheet instanton sector.
These contributions are of the form~\cite{Anderson, Cicoli}
\begin{align}
W^{\mathrm{worldsheet}}_{\mathrm{np}} = B e^{-n_i T^i} ~. \label{eq:non-pert-2}
\end{align}
where $B$ and $n^i$ are numbers.
Here, we choose an integral basis $\{T^i\}$ for the K\"ahler moduli space.
As a result $n_i$ are integers which count the number of times string worldsheets wrap holomorphically around two-cycles on the Calabi--Yau.
The values of $n_i$ can be determined through modular invariance and anomaly constraints on the superpotential.

We seek to provide a mechanism that can apply to large classes of Calabi--Yau compactifications.
As such, in order to maintain generality, we do not compute the values of parameters such as $\beta_i$ and $n_i$ explicitly, though this can be done by selecting a specific model and working through the details.

\subsection*{Kinetic Terms}
The kinetic terms for the $\tau$ axions are:
\begin{align}
\mathcal{L}_{\mathrm{kinetic}} = -K_{i\bar{j}}\partial_\mu \tau_i \partial^\mu \tau_{\bar j} ~,
\end{align}
where $K_{i\bar{j}}$ is computed from (\ref{eq:Kahler}) in the standard way and is given by
\begin{align}\label{eq:kinetic-matrix}
K_{i\bar{j}}=\frac{3}{4(t^3_1-t_2^3)^2}\left(\begin{array}{cc}
t^4_1 + 2 t_1 t^3_2 & -3 t_1^2 t_2^2\\
-3 t_1^2 t_2^2 & t^4_2 + 2 t_2 t^3_1
\end{array}\right).
\end{align}
We see that there is kinetic mixing between the two axions, and so we shall need to diagonalize and canonically normalize the kinetic terms of the fields.

\subsection*{Scalar Potential}
As usual the F-term contribution to the scalar potential is given as follows:
\begin{align}
V_F= e^K \left( K^{i\bar{j}} D_i W \bar D_{\bar j} \bar W -3 |W|^2 \right) ,
\end{align}
where $D_i W=W,_i+K,_iW$ and $K^{i\bar j}$ is the inverse of the K\"ahler metric on the K\"ahler structure moduli space.
Given the form of the superpotential and the fact that the K\"ahler sector satisfies the no-scale condition and assuming $W_p$ is large compared to superpotential terms (\ref{eq:non-pert-1}) and (\ref{eq:non-pert-2}) coming from the non-perturbative sector, a computation similar to the one in~\cite{long} yields the following scalar potential:
\begin{align}
V_F=  e^K \left\{K^{i\bar{j}} W_p K_i \partial_{\bar{j}}\overline{W}_{\mathrm{np}}+\mathrm{c.c.}\right\} ~, \label{eq:potential}
\end{align}
where 
$W_{\mathrm{np}}$ is the sum of the non-perturbative superpotentials.
Since we have assumed that $s$ is fixed to $s^*$, the prefactor in the potential (\ref{eq:potential}) becomes
\begin{align}
e^K= \frac{1}{16 s^* (t^3_1-t^3_2)} ~,
\end{align}
where we have used the the explicit expression for $\widetilde{\K}$ is given by
\begin{align}
\widetilde{\K}= t_1^3-t_2^3 ~.
\end{align}
$\widetilde{\K}$ is the volume of the internal Calabi--Yau up to an order one constant.
Consistency requires that $\widetilde{\K}\gg 1$ and that the matrix $\frac{\partial^2 \widetilde{\K}}{\partial t_i\partial t_j}$ has signature $(1,-1)$~\cite{Candelas}.
Furthermore,  $s^*>1$ is required for string theory to be in the weakly coupling regime.
In the numerical example that we present in Section~\ref{sec:example} we have been careful to pick positive values of $t_1$ and $t_2$ as well as $s^*>1$ so that these requirements are satisfied.

We require a large volume limit of the Calabi--Yau in order to ensure that certain effects are perturbative.
We are agnostic as to how large the volume should be so long as there is a separation of scale between the Planck length and the string length.

With these ingredients it is straightforward to compute the scalar potential.
However, in order to implement the decay constant misalignment mechanism we need to add to this scalar potential a supersymmetry breaking uplifting term $V_\mathrm{up}$ whose precise form is computed in Appendix~\ref{sec:scalar-potential}.
We shall assume the value of $s^*$ is chosen so that our gauge coupling constants remain in the perturbative regime, and that $\sigma$ is stabilized at $\frac{2n \pi}{\alpha}$, where $n$ is an integer.
Then our non-perturbative superpotential takes the form
\begin{align}
W_{\mathrm{np}}= A e^{\alpha \beta_i T^i} + B e^{-n_i T^i} ~, \label{eq:non-perturbative-superpotential}
\end{align} 
with
\begin{align}
A= A' e^{-\alpha s^*} ~.
\end{align}

The full scalar potential for the construction is then given by
\begin{align}
V&=V_F+ V_{\mathrm{up}}\nonumber\\
&= \Lambda^4_A \left[1-\cos{\{\alpha(\beta_1\tau_1+\beta_2\tau_2)\}}\right] + \Lambda^4_B \left[1-\cos{(n_1\tau_1+n_2\tau_2)}\right] \label{eq:final_potential1} ~,
\end{align}
where the two $\Lambda$s are:
\begin{align}
\Lambda_A^4=\frac{ A |W_p| \alpha (\beta_1 t_1+\beta_2 t_2)e^{\alpha (\beta_1 t_1+\beta_2 t_2)   } } { 4 s^*(t_1^3-t_2^3)} ~, \quad
\Lambda_B^4=-\frac{B |W_p|  (n_1 t_1+n_2 t_2)e^{-( n_1 t_1+n_2 t_2 )   }} {4 s^* (t_1^3-t_2^3)}  ~.\end{align}
The uplifting term $V_{\mathrm{up}}$ in the above equation is given by (see Appendix~\ref{sec:scalar-potential})
\begin{align}
V_{\mathrm{up}}= 2 e^K (\mathcal{M}'+\mathcal{N}') ~, \label{eq:uplift}
\end{align}
where we have defined
\begin{align}
\mathcal{M}' &= 2 e^{\alpha\beta_i t^i} A \alpha t_i\beta_i W_p ~,\nonumber\\
\mathcal{N}' &= -2 e^{-n_i t^i} B t_i n_i W_p ~.
\end{align}
The source of the uplifting term will be explained below.
For (\ref{eq:final_potential1}),  in terms of the canonically normalized fields, the determinant of the Hessian of the potential at the minimum is given by
\begin{align}
\text{det} \ \partial_i \partial_j V |_{\tau=0}\ &= -\frac{ \alpha^2  \mathcal{M'} \mathcal{N'} (n_2 \beta_1-n_1\beta_2)^2}{72 s^*{^2} t_1 t_2} ~.
\end{align}

\subsection*{The Uplifting Term}
The uplifting term $V_{\mathrm{up}}$ in the type IIB setting generically comes from anti-D-branes that break the supersymmetry.
In a heterotic setting, since there are no D-branes it is customary to invoke the D-terms associated with the anomalous $U(1)$ factors~\cite{Burgess} that are generic to heterotic compactifications to do the same job.
Fortunately in the setting of~\cite{Anderson} there are anomalous $U(1)$ line bundles that supply us with such D-terms.
According to~\cite{Anderson}, the D-term associated with the $I$-th line bundle is
\begin{align}
D_I=\frac{c^i_1(\mathcal{L}_I) d_{ijk} t^j t^k }{d_{ijk} t^i t^j t^k} =\frac{  \left (c^1_1(\mathcal{L}_I) t_1^2- c^2_1(\mathcal{L}_I) t_2^2 \right)}{(t_1^3-t_2^3)} ~,
\end{align}
where $c^i_1(\mathcal{L}_I)$ is related to the first Chern class $c_1(\mathcal{L}_I)$ of the $I$-th holomorphic line bundle in the following way:
\begin{align}
c_1(\mathcal{L}_I) = c^i_1(\mathcal{L}_I) \omega_i ~.
\end{align}
The $\omega_i$ are the basis for the second cohomology group of the Calabi--Yau manifold $X$.
The D-terms contributes the following term to the potential
\begin{align}
V_D = \frac{1}{2} \sum_I D_I D_I ~,
\end{align}
which must be tuned to give the uplifting term (\ref{eq:uplift}).
It should be noted that whether this proposed $D$-term is able to act as $V_{\mathrm{up}}$ is a model dependent statement and should be examined on a case by case basis.
\subsection*{Large Effective Decay Constant from Near Alignment} 
The final step is to canonically normalize the axions fields $\tau_1$ and $\tau_2$ before comparing the final potential with (\ref{eq:original_potential}).
Since the matrix of coefficients for the kinetic terms (\ref{eq:kinetic-matrix}) is field dependent, it is rather awkward to do so.
In the next section, the axions are canonically normalized numerically after choosing generic values of the fields $t_1$ and $t_2$.
(See, for example,~\cite{Svrcek:2006yi, Baumann:2014nda}.)
We denote the canonically normalized axions by $\psi_1$ and $\psi_2$.
The canonically normalized fields $\psi_i$ when expressed in terms of the axions $\tau_i$ involve factors coming from $K_{i\bar{j}}$.
From~\eref{eq:kinetic-matrix}, we observe that the matrix scales like $1/(a_{ij} t^i t^j)$.
In the large volume limit, this is a small number.
The fundamental axion decay constant is therefore decreased and is always sub-Planckian.
This shows why it is so difficult to produce large fundamental axion decay constants while staying within a perturbative large volume regime.
The crucial point of the KNP mechanism is that the effective decay constant, which controls the dynamics of inflation, is nevertheless large.
The examples we consider are constructed to illustrate this structure.

The examples should not be fine tuned.
In a natural theory, we expect all dimensionless parameters to be of order one in string units.
Implicit in the construction is the assumption that the moduli fields and the values of $\alpha, \beta_i, n_i, W_p, A', B$, and $s^*$ can be chosen to lie within one or two orders of magnitude of unity.
(The one or two orders of magnitude accounts for factors such as $8\pi^2$ that appear in the one-loop beta function coefficient $\alpha$.)
The precise determination of all of these factors is a subtle and non-trivial question (see, for example,~\cite{Buchbinder}) which should be addressed within the context of a specific model and lies outside the scope of this work.
Experimentally, as we indicate in the next section, we do not find it difficult to select values of the parameters that satisfy naturalness.


\subsection*{Misalignment}
We now consider an analytical example for how one can implement in the potential (\ref{eq:final_potential1}) a tiny misalignment to give rise to a large effective decay constant.
This particular example echoes the analytical example given~\cite{long} even though the physical origin of our parameters is very different from theirs.

Given that we have
\begin{align}
&\Lambda^4_A = 2 e^K \mathcal{M}' \;\;\;\mathrm{and}\;\;\;\Lambda^4_B = 2 e^K \mathcal{N}'
\end{align}
we can choose
\begin{align}
 n_1=n_2=\alpha \beta_1 \equiv n,  \;\; \alpha \beta_2=\frac{\alpha \beta_1}{1+\delta}, \;\; \Lambda_B^4=\Lambda^4 ,\;\mathrm{and}\;\; \Lambda_A^4=\delta^p\Lambda^4,
\end{align}
where $0<\delta\ll 1$ and $p>0$ in order to break the alignment slightly.
Let us denote the mass eigenstates by $\varphi_1$ and $\varphi_2$ with the corresponding mass eigenvalues: 
\begin{align}
m_1^2= \delta^{2+p} \ \frac{\Lambda^4 n^2}{ 2 }\  [1+\mathcal{O} (\delta)], \;\ m_2^2= 2 \Lambda^4 n^2 \ [1+\mathcal{O} (\delta)].
\end{align}
In terms of the mass eigenstates the potential can be rewritten as
\begin{align}
V= \Lambda^4\left[1-\text{cos} \left(\sqrt{2}  n \varphi_2 + \frac{\delta^{p+1} n \varphi_1   }{2 \sqrt{2}  }  \right)      \right] +\delta^p \Lambda^4\left[     1-\text{cos} \left(-\sqrt{2}  n \varphi_2 + \frac{\delta n \varphi_1   }{ \sqrt{2}  }  \right)      \right].
\end{align}
The first term of this potential serves to fix $\varphi_2$ to zero while the second term gives rise to and effective decay constant for $\varphi_1$ whose value is given by
\begin{align}
f_{\text{eff}}\sim\frac{1}{n~\delta}~.
\end{align}
As mentioned above the proportionality constant in the above expression also involves a factor arising from canonically normalizing the fields.
In the next section, where we choose explicit  numbers to generates a large effective decay constant the fields, $\psi_1$ and $\psi_2$, are canonically normalized.
\section{Example }\label{sec:example}
In this section, we will generate an example for the near alignment mechanism.
This should be considered a toy model with a discussion about how this could be implemented in a realistic Calabi--Yau manifold postponed to a future work.
\begin{figure}[htbp]
\begin{center}
\includegraphics[scale=1.0]{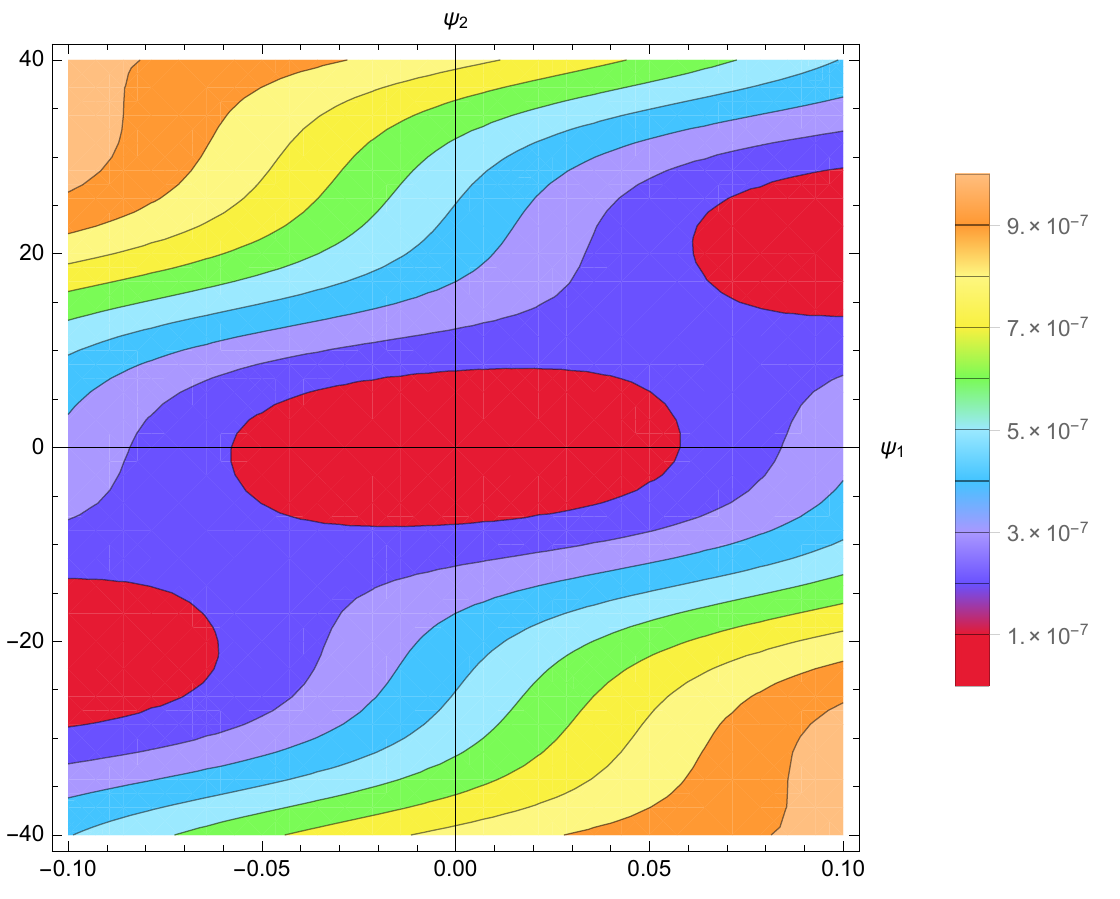}
\caption{Contour plot for the two axions scalar potential.
Inflation is along the $\psi_2$ direction.}
\label{fig:Contour}
\end{center}
\end{figure}
For the following values
\begin{align} 
\alpha&=0.105 ~, \quad \beta_1=18 ~, \quad \beta_2=37 ~, \quad n_1=1 ~, \quad n_2=2 ~, \quad W_p=0.3 ~,\nonumber \\
A'&=0.1 ~, \quad B=-20 ~, \quad t_1=5 ~, \quad t_2=1 ~, \quad s^*=168 ~,
\end{align}
 we get the following mass eigenvalues:
\begin{align}
m_1= 7.9 \times 10^{-3} \ M_{\mathrm{Pl}} ~,\;\
m_2= 6.0\times 10^{-5} \ M_{\mathrm{Pl}} ~.
\end{align}
We have been careful to choose numbers that are close to order one.
The potential in terms of the canonically normalized mass eigenstates can be written as:
\begin{align}
\frac{V}{M_{\mathrm{Pl}}^4} &= 5.3\times 10^{-7}- (4.6\times 10^{-7})\; \cos{\left[  \frac{9.44 \psi_1}{M_{\mathrm{Pl}}} -\frac{0.05\psi_2}{M_{\mathrm{Pl}}}     \right ]}\nn\\
&-(6.54\times 10^{-8})\; \text{cos} \left[  \frac{18.13\psi_1}{M_{\mathrm{Pl}}} +\frac{0.19\psi_2}{M_{\mathrm{Pl}}}     \right ] ~.
\end{align}
The hierarchy between the decay constants allows us to integrate out the $\psi_1$ field and we are then left with $\psi_2$.
Identifying the remaining $\Lambda^4 \approx 6.54\times 10^{-8}$ (\emph{cf.}\ (\ref{eq:one-field})), the coefficient of $\psi^2_2$ yields
an effective decay constant of roughly $f_{\mathrm{eff}}\approx 11.3~M_\mathrm{Pl}$.
Moreover, for these choices of parameters the non-perturbative superpotential is much smaller than the classical superpotential.

We have found that it is fairly easy to obtain other explicit examples which yield large values of effective decay constants for a range of parameters that lie within several orders of magnitude of unity.
This indicates that fine tuning of parameters is not necessary to implement the KNP mechanism in the heterotic context.

\section{Discussion}\label{sec:disc}
In this paper, we have proposed an elementary mechanism by which large field inflation can be realized in heterotic string theory.
Our construction is inspired by that of~\cite{long}.
Like them, we make use of the scheme of natural inflation~\cite{Freese} and the decay constant alignment mechanism proposed by~\cite{kim}.
We do not address the problem of moduli stabilization explicitly in this paper, but propose that this issue can be addressed using the techniques developed in~\cite{Anderson, Anderson2, Anderson3}.
While most string cosmology work in recent years has been performed in a type II setting, the heterotic setting described in our paper seems particularly simple and natural.
Similar work has been recently done in~\cite{Ben-Dayan}.

The main ingredients of our construction are as follows: a classical superpotential that only depends on the complex structure moduli and non-perturbative superpotentials coming from gaugino condensation, and string instantons for the the axions arising from the K\"ahler structure moduli.

These axions appear with two different linear combinations in the scalar potential that results from the interplay of these superpotential terms.
When we tune the model so that these two linear combinations are slightly different, we can break the alignment and can generate a large value for the effective decay constant.
In our setting we achieve uplifting of the vacuum using D-terms proposed in~\cite{Anderson}.
We then demonstrate the effectiveness of our construction with a numerical example by choosing plausible values of the tuneable parameters that contribute to the model.

In this way, we are able to generate a large enough effective decay constant that realizes cosmological inflation.
For the simple numerical example that we have presented, the effective decay constant is larger than $10 M_\mathrm{Pl}$ and $\Lambda$ is greater than the Hubble constant.
If the BICEP2 observation of the tensor-to-scalar ratio $r$ turns out to be correct, then these numbers seem consistent with the inflationary models supported with that data~\cite{Ade}.

Our analysis has been at tree level in $g_s$.
An important issue for this type of construction is whether the $g_s$ threshold corrections will spoil the flatness of the potential.
In~\cite{Cicoli}, the authors argued that this threshold effect is actually subdominant with respect to the $\alpha'$ ones for large volume, so it may be reasonable to neglect their effect.
As a future project, we believe that it would be an interesting to exploit the database of Calabi--Yau threefolds~\cite{Kreuzer:2000xy, Altman:2014bfa} to look for explicit examples of manifolds which would realize our construction in more concrete settings.

\textbf{Note:} As this work was in the final stages a preprint~\cite{ako} appeared on arxiv which also implements natural inflation from heterotic strings.
However, their construction is very different from ours as they did not try to implement the KNP proposal from axions with sub-Planckian decay constants.
\section*{Acknowledgements} 
We are grateful to Liam McAllister for commenting on a draft of this paper.
We would also like to thank James Gray and Suresh Nampuri for helpful conversations.
The research of SSH and VJ is supported by the South African Research Chairs Initiative of the Department of Science and Technology and the National Research Foundation.
SSH would also like to thank Perimeter Institute for their hospitality where part of the work was done.
Research at Perimeter Institute is supported by the Government of Canada through Industry Canada and by the Province of Ontario through the Ministry of Economic Development and Innovation.
VJ is grateful to the string group at Queen Mary, University of London for kindly hosting him during the concluding stages of this work.
TA and SSH are grateful to the organizers of the Summer Workshop at the Simons Center for Geometry and Physics for their hospitality where part of this work was done.
\appendix
\section{The Scalar Potential} \label{sec:scalar-potential}
In a supersymmetric theory the holomorphic superpotential $W$ gives rise to the scalar potential via 
\begin{align}
V_F= e^K \left( K^{I\bar{J}} D_I W \bar D_{\bar J} \bar W -3 |W|^2 \right) ~.
\end{align}
As explained in the main text, the superpotential for our model has the form:
\begin{align}
W=W_p + W_{\mathrm{np}} ~,
\end{align}
where we assume that the classical superpotential $W_p$ only depends on the complex structure moduli whose values have been fixed to yield a small $W_p$ using a mechanism similar to~\cite{Anderson}.
Additionally, we assume that the axiodilaton $S$ has been fixed by imposing its  $F$-term equation.
Thus, the expression for the potential in terms of the superfields now only contains derivatives with respect to the K\"ahler moduli.
Furthermore, by using the identity $K^{i\bar{j}} K_i K_{\bar{j}}=3$, we arrive at
\begin{align}
e^{-K} V_F = K^{i\bar{j}} W_p K_{i} \partial_{\bar{j}}\overline{W}_{\mathrm{np}}+\mathrm{c.c.} ~,
\end{align}
where we have made the additional assumption that $W_{\mathrm{np}}$ is smaller than $W_p$ so that we can ignore terms which are quadratic in $W_{\mathrm{np}}$.
A key observation here is that, as pointed out in (\ref{eq:0th-superpotential}), $W_p$ is function of only the complex structure moduli $Z^a$, whereas, after having fixed $S$, the non-perturbative superpotential (\ref{eq:non-perturbative-superpotential}) is exclusively a function of the K\"ahler structure moduli fields.

From the expression of the non-perturbative superpotential we have:
\begin{align}
\partial_{\bar{j}} \overline{W}_{\mathrm{np}}= A\alpha \beta_{j} \;e^{\alpha\beta_k\bar{T}^k} - B n_j e^{-n_k \bar{T}^k}.
\end{align}
Noting that
\begin{align}
K^{i\bar{j}} K_i =-2 t_j
\end{align}
as well as
\begin{align}
M &= \alpha \beta_{i}\bar{T}^i\nonumber ~, \\
N & = -n_i \bar{T}^i ~,
\end{align}
we have
\begin{align}
e^{-K}V_F &=2 t_i( - e^M A \alpha \beta_i + e^N B n_i )W_p + \mathrm{c.c.}\\
&\equiv  e^M \mathcal{M} + e^N \mathcal{N} + \mathrm{c.c.}
\end{align}
Next, using the decomposition (\ref{eq:Kahler-fields}), we isolate the part of the potential that is a function of $\t_1$ and $\t_2$:
\begin{align}
e^{-K}V_F &=- e^{-i\alpha (\beta_1 \tau_1+\beta_2\tau_2)} \mathcal{M}' - e^{i (n_1\tau_1+ n_2\tau_2)} \mathcal{N}' + \mathrm{c.c.} ~,
\end{align}
where we have defined
\begin{align}
\mathcal{M}' &= -e^{\alpha\beta_i t^i}\mathcal{M} ~,\nonumber\\
\mathcal{N}' &= -e^{-n_i t^i} \mathcal{N} ~.
\end{align}
We now assume that the functions $\mathcal{M}'$ and $\mathcal{N}'$ can be chosen to be real  so that the potential can now be expressed as
\begin{align}
V = e^K\left\{ 2 \mathcal{M}' \left[1-\cos{\alpha(\beta_1\tau_1+\beta_2\tau_2)}\right] + 2\mathcal{N}'\left[1-\cos{(n_1\tau_1+n_2\tau_2)}\right] \right\} ~,
\end{align}
where we have set:
\begin{align}
V_{\mathrm{up}}= 2 e^K (\mathcal{M}'+\mathcal{N}') ~.
\end{align}

\bibliography{references}

\bibliographystyle{utphysmodb}

\end{document}